\documentclass[superscriptaddress,amsmath,amssymb,aps,pra]{revtex4-2}

\usepackage{graphicx}% Include figure files
\usepackage{dcolumn}% Align table columns on decimal point
\usepackage{bm}% bold math
\usepackage{amsmath} 
\usepackage[colorlinks]{hyperref}% add hypertext capabilities
\usepackage{caption}
\usepackage{subcaption}
\usepackage{cancel}
\usepackage{gensymb}
\usepackage{float}
\usepackage{svg}
\usepackage[english]{babel}

\begin{document}

\title{State Engineering via Nonlinear Interferometry with Linear Spectral Phases}

\author{Cody Charles Payne\textsuperscript{*}}
\affiliation{Department of Physics and Astrophysics, University of North Dakota, Grand Forks, North Dakota, USA}
\email{cody.payne@und.edu}
\author{Elaganuru Bashaiah}
\affiliation{Department of Physics and Astrophysics, University of North Dakota, Grand Forks, North Dakota, USA}
\author{Markus Allgaier}
\affiliation{Department of Physics and Astrophysics, University of North Dakota, Grand Forks, North Dakota, USA}

\begin{abstract}
Many protocols within quantum cryptography, communications, and computing require the ability to generate entangled states as well as spectral qudits. Nonlinear interferometry is a viable way to engineer these complex quantum states of light. However, it is difficult to achieve a high level of control over spectral correlations. Here, we present a protocol utilizing a nonlinear interferometer with linear spectral phases that can generate both high-dimensional spectral qudits and high-dimensional entangled states. We model the effect of loss and loss of overlap on interference visibility and thereby on the states generated.
\end{abstract}

\maketitle

\section{Introduction}
Many of the challenges in developing practical and effective quantum technologies concern the creation of physical qubits and the tailoring of non-trivial quantum states. These two tasks fall under the umbrella of quantum state engineering. Engineering the spectral correlations between single photon pairs produced through spontaneous parametric down conversion (SPDC) can be an effective way of accomplishing state engineering of this sort.

Often, one wishes to obtain a single photon in a pure spectral state by heralding one photon of the down-converted pair \cite{spectral_multiplexing_5}. Such pure state single photons can then be used in quantum protocols utilizing Hong-Ou-Mandel (HOM) interference \cite{HOM}. Spectral state engineering can also be turned toward developing non-trivial quantum states on the spectral degree of freedom (DoF) directly, thus generating states suitable for frequency-space quantum protocols \cite{PhysRevLett.117.223601,Lukens:16,Lukens:17,PhysRevLett.130.200602}. By taking advantage of the spectral modes of photons, quantum protocols can in principle be conducted utilizing a much larger `alphabet' than is afforded by path and polarization DoF's. To implement high-dimensional quantum protocols, it is useful to have access to two classes of quantum states: high-dimensional spectrally entangled states and single-photon spectral superposition states.

High-dimensional entanglement (HDE) in the basis of individually pure joint spectral modes has been accomplished by controlled frequency conversion of one photon from the pair \cite{spectral_multiplexing_3,spectral_multiplexing_4}, domain engineering to directly control the structure of the phase matching \cite{frequency_entanglement,Shukhin:24} direct control of the pump spectrum \cite{spectral_multiplexing_6}, use of a Fabry-P\'{e}rot cavity \cite{Cheng2023}, and by nonlinear interferometry (NLI) \cite{PhysRevA.102.033718,Li:19,Su:19,article,Ferreri2021spectrallymultimode,PhysRevA.97.053827}. Meanwhile, single-photon spectral superpositions have been achieved by frequency conversion using Bragg scattering four-wave mixing \cite{PhysRevLett.117.223601,Lee2024} and frequency-conversion by difference frequency generation \cite{Aguayo-Alvarado2022}. In the context of three- and four-wave mixing,  Shukhin et al. \cite{Shukhin:24} discussed the fact that frequency-resolving postselection on a joint spectral ``grid state'' can yield spectral superpositions. If the grid state is composed of a series of separable, pure spectral modes, then detection of the heralding photon in a particular frequency bin will project its sibling photon onto a superposition of the spectral modes that lie on the corresponding row or column of the lattice. Schemes for the production of such a grid state have been developed utilizing domain engineering \cite{Hurvitz:23}, domain engineering combined with the application of a time delay to the pump spectrum \cite{Shukhin:24}, and Fabry-P\'{e}rot cavity modulation \cite{PhysRevA.102.012607}. 

Of the schemes discussed above, NLI is particularly desirable. Briefly, an NLI consists of a series of identical nonlinear crystals in any of which a pair generation process can occur. It is highly unlikely that a pair generation process will occur in two or more of the crystals simultaneously, thus for any given photon pair, the indistinguishability as to which crystal produced that pair leads to an interference effect. This interference can be controlled by the introduction of a spectral phase between each of the crystals, such that the joint spectrum of the photon pair is modulated in a way that depends on the form of that phase.  Of the advantages offered by this approach, NLI schemes can be implemented using only off-the-shelf optical components, and thus the careful fabrication processes needed for schemes involving domain engineering or Fabry-P\'{e}rot cavities is not needed. Furthermore, by adjusting the relative phase between the pair creation events, the parameters of the joint spectral states engineered by utilizing an NLI can be reconfigured.
NLIs have been implemented different numbers of crystals and mechanisms to apply phases between them, including dispersion and birefringence. However, none of these achieve the fully arbitrary modulation of the joint spectrum which we seek to demonstrate here \cite{PhysRevA.102.033718,Li:19, article, Su:19, Ferreri2021spectrallymultimode}. 

Here we present a scheme that consists of four nonlinear crystals in series, where phases are applied in the form of time delays to pump, signal, and idler photons independently. We thus find a reconfigurable scheme that, depending on the choice of time delays, can either produce an HDE state with identical spectral modes or a grid state upon which postselection can be done to yield spectral qudit states as already discussed. We will also analyze the effect of loss on the modal purity of the states produced via NLI modulation. The effects of loss and crystal dispersion are relevant in any realistic implementation of this scheme as it affects the interference visibility in non-trivial ways, but it is not often discussed in previous treatments of the subject. We thus achieve a scheme that is more versatile than previous schemes in that it can generate both HDE and grid states with only linear spectral phases, at the cost of optical elements to separate and recombine pump, signal, and idler photons at every crystal. In the case of the HDE states, which are the typical outcome of previous schemes, we generate entanglement in a basis consisting of spectral modes that are approximately Gaussian envelopes. The tradeoff in this case is that the scheme requires type-II phase matching, which cannot yield a joint spectral intensity entirely along the $-45^{\circ}$ axis. This will turn out to slightly diminish the quality of the HDE state. 

This paper is organized as follows: First, we give a brief overview of the theory underlying analysis of the biphoton joint spectrum and how this spectrum is modulated by the introduction of spectral phases in an NLI. We then present the scheme, outlining the particular sequences of time delays which give rise to the desired modulation and deriving how the modulation function arises from these sequences. We then present the results of simulations of these schemes and brief discussion of how they may be implemented in a laboratory setting. Finally, we revisit the previous simulations while introducing the effects of loss and analyze how the spectral mode decomposition of the states generated by the NLI is affected by this loss.

\section{Theory}

The interaction Hamiltonian for collinear spontaneous parametric down conversion involving a pump photon with spectral profile $\alpha(\omega_p)$ propagating in the $z$ direction in a crystal of length $L$ and with coupling amplitude $g$ is given by:

\begin{align}
    &\hat{H}_{\text{SPDC}}(t)=\nonumber\\
    &\hbar g\int_{-\frac{L}{2}}^{\frac{L}{2}}\mathrm{d}z\int\mathrm{d}\omega_s\mathrm{d}\omega_i\mathrm{d}\omega_p\,\alpha(\omega_p)\,\beta(\omega_p,\omega_s,\omega_i)\,\,e^{i(\omega_p-\omega_s-\omega_i)t}\,e^{i(k_s+k_i-k_p)z}\,\hat{a}^{\dagger}_{\omega_s}\hat{a}^{\dagger}_{\omega_i}\hat{a}_{\omega_p}\nonumber\\
    &=\hbar\Omega\int\mathrm{d}\omega_s\mathrm{d}\omega_i\mathrm{d}\omega_p\,\alpha(\omega_p)\,\beta(\omega_p,\omega_s,\omega_i)\,\mathrm{sinc}\left(\frac{\Delta k L}{2}\right)\,e^{i(\omega_p-\omega_s-\omega_i) t}\,e^{i\frac{\Delta k L}{2}}\,\hat{a}^{\dagger}_{\omega_s}\hat{a}^{\dagger}_{\omega_i}\hat{a}_{\omega_p},
\end{align}

\noindent where $\beta\equiv\beta_p\,\beta_s\,\beta_i$ is a modulation term which encodes phase factors acquired by pump, signal, and idler photons, $\Omega={igL}$, and $\Delta k=k_s+k_i-k_p$ and where $\hat{a}_{\omega_j}\,(\hat{a}^{\dagger}_{\omega_j})$ is the annihilation (creation) operator for a photon of angular frequency $\omega_j$, respectively. To obtain the corresponding unitary $\hat{U}=\hat{\mathcal{T}}e^{-\frac{i}{\hbar}\int_{t_0}^{t_f}\mathrm{d}t\,\hat{H}(t)}$, we need to calculate the time integral:

\begin{align}
    \int_{t_0}^{t_f}\mathrm{d}t\,\hat{H}_{\text{SPDC}}(t)&=\int_{-L/2c}^{L/2c}\mathrm{d}t\,\hat{H}_{\text{SPDC}}(t)\nonumber\\
    &=\hbar\kappa\int\mathrm{d}\omega_p\mathrm{d}\omega_s\mathrm{d}\omega_i\,\alpha(\omega_p)\,\beta(\omega_p,\omega_s,\omega_i)\,\mathrm{sinc}\left(\frac{\Delta k L}{2}\right)\,\nonumber\\
    &\quad\times\mathrm{sinc}\left(\frac{(\omega_p-\omega_s-\omega_i)L}{2c}\right)e^{i\frac{\Delta k L}{2}}e^{i\frac{\Delta\omega L}{2c}}\,\hat{a}^{\dagger}_{\omega_s}\hat{a}^{\dagger}_{\omega_i}\hat{a}_{\omega_p},
\end{align}

\noindent where $\kappa=\frac{i\Omega L}{c}$. When the crystal length $L$ is sufficiently long, the $\mathrm{sinc}$ function can be approximated as a Dirac delta. More precisely, this approximation can be made when the interaction time $\Delta t$, which is on the order of $L/c$, is large in comparison to an optical period $T_j=2\pi/\omega_{j}$, thus:

\begin{align}\label{time_integral}
\int_{t_0}^{t_f}&\mathrm{d}t\,\hat{H}_{\text{SPDC}}(t)\approx
\quad\hbar\kappa\int\mathrm{d}\omega_p\mathrm{d}\omega_s\mathrm{d}\omega_i\,\alpha(\omega_p)\,\beta(\omega_p,\omega_s,\omega_i)\,\mathrm{sinc}\left(\frac{\Delta k L}{2}\right) \nonumber\\ &\quad\times\delta\left({\omega_p-\omega_s-\omega_i}\right)e^{i\frac{\Delta k L}{2}}e^{i\frac{\Delta\omega L}{2c}}\,\hat{a}^{\dagger}_{\omega_s}\hat{a}^{\dagger}_{\omega_i}\hat{a}_{\omega_p} \nonumber \\
&=\hbar\kappa\int\mathrm{d}\omega_s\mathrm{d}\omega_i\,\alpha(\omega_s+\omega_i)\,\beta(\omega_s+\omega_i,\omega_s,\omega_i)\,\mathrm{sinc}\left(\frac{\Delta k L}{2}\right)\,e^{i\frac{\Delta k L}{2}}\,\hat{a}^{\dagger}_{\omega_s}\hat{a}^{\dagger}_{\omega_i}\hat{a}_{\omega_s+\omega_i}.
\end{align}

Note that this particular derivation was performed in large part by following the broad outline of a similar derivation performed by Brecht \cite{Brecht2014}. 

The kernel of the resultant integral of \eqref{time_integral} is known as the joint spectral amplitude (JSA) which dictates how the amplitudes of the spectrum of the daughter photons are distributed. The unitary for time evolution involving a nonlinear crystal in which SPDC can take place is then:

\begin{align}
    \hat{U}_{\text{SPDC}}&=\hat{\mathcal{T}}e^{-\frac{i}{\hbar}\int_{t_0}^{t_f}\mathrm{d}t\,\hat{H}_{\text{SPDC}}(t)}\nonumber\\
    &=1-i\kappa\int\mathrm{d}\omega_s\mathrm{d}\omega_i\,JSA(\omega_s,\omega_i)\,\hat{a}^{\dagger}_{\omega_s}\hat{a}^{\dagger}_{\omega_i}\hat{a}_{\omega_s+\omega_i}\nonumber\\
    &\qquad-\frac{\kappa^2\hat{\mathcal{T}}}{2}\,\left(\int\mathrm{d}\omega_s\mathrm{d}\omega_i\,JSA(\omega_s,\omega_i)\,\hat{a}^{\dagger}_{\omega_s}\hat{a}^{\dagger}_{\omega_i}\hat{a}_{\omega_s+\omega_i}\right)^2+\ldots\nonumber\\
    &=\hat{U}_{\text{SPDC}}^{(0)}-i\kappa\,\hat{U}_{\text{SPDC}}^{(1)}-\frac{\kappa^2}{2}\,\hat{U}_{\text{SPDC}}^{(2)}+\ldots.
\end{align}

When discussing SPDC, it is customary to only consider the first order unitary which yields exactly one pair of daughter photons given one pump photon, i.e.:

\begin{equation}
    \hat{U}^{(1)}_{\text{SPDC}}|\omega_p,0_s,0_i\rangle=|0_p\rangle\otimes\int\mathrm{d}\omega_s\mathrm{d}\omega_i\,JSA(\omega_s,\omega_i)\,\hat{a}^{\dagger}_{\omega_s}\hat{a}^{\dagger}_{\omega_i}|0_s,0_i\rangle\equiv|0_p\rangle\otimes|JSA\rangle
.\end{equation}

In general, the state $|JSA\rangle$ is a non-separable mixed state. The JSA shall be unaffected by detector effects, which are assumed to have flat temporal and spectral response. However, it is possible via singular value decomposition to find a minimal set of spectral modes $\{\psi_k\}$ and $\{\phi_k\}$ such that:

\begin{equation}\label{mode_eqn}
    |JSA\rangle=\sum_kc_k\int\mathrm{d}\omega_s\mathrm{d}\omega_i\,\psi_k(\omega_s)\phi_k(\omega_i)\,\hat{a}_{\omega_s}^\dagger\hat{a}_{\omega_i}^{\dagger}|\text{vac}\rangle.
\end{equation}

The spectral purity can then be quantified by the Schmidt number denoted $K$ which is defined in terms of the coefficients $\{c_k\}$ from \eqref{mode_eqn} as:

\begin{equation}
    K=\frac{1}{\sum_k|c_k|^4}=\frac{1}{\mathrm{Tr}\left(\hat{\rho}^2\right)},
\end{equation}

\noindent where $\hat{\rho}$ is the density matrix of the joint spectral state.

Let the term $JSA_0$, defined as:

\begin{equation}
    JSA_0=\alpha(\omega_s+\omega_i)\,\mathrm{sinc}\left(\frac{\Delta k L}{2}\right)\,e^{i\frac{\Delta k L}{2}}
\end{equation}

\noindent denote the  joint spectral amplitude for a single crystal, which is the joint spectrum that would arise in the absence of the modulation term $\beta(\omega_s,\omega_i)$. The general modulated joint spectral state is then given by:

\begin{equation}
JSA(\omega_s,\omega_i)=\beta(\omega_s,\omega_i)\times JSA_0(\omega_s,\omega_i).
\end{equation}

In the case of an NLI consisting of a series of $N$ crystals, the form of $JSA_0$ arises from the structure of the nonlinear crystals individually, and thus is the same 
for all the crystals in the series provided that they are mutually identical. The modulation then arises from choosing the factors $\beta^{(\mu)}$ appropriately for each crystal (where $\mu$ denotes the crystal index). By summing up the amplitudes associated with the pair production in each crystal, we obtain the total joint spectral amplitude given by:

\begin{equation}
JSA(\omega_s,\omega_i)=JSA_0(\omega_s,\omega_i)\,\sum_{\mu=1}^N\beta^{(\mu)}=\beta(\omega_s,\omega_i)\,JSA_0(\omega_s,\omega_i),
\end{equation}

\noindent where:

\begin{equation}
    \beta^{(\mu)}=\prod_{j=p,s,i}\beta_j^{(\mu)}.
\end{equation}

The index $\mu$ corresponds to the the crystal and the fields (pump, signal and idler) involved in the PDC process taking place in that crystal. To find $\beta_j^{(\mu)}$, we note first that for our purposes here, we will consider type-II phase matching such that the pump, signal, and idler photons can be separated by polarization and dichroic optics and then manipulated individually. In this case, it is possible to choose materials of length $\ell_{j}^{(\mu)}$ and refractive indices $n_j^{(\mu)}(\omega_j)$ for each crystal indexed $\mu$ and each photon indexed $j$. Then we can define a phase $\varphi_j^{(\mu)}(\omega_j)$ specific to each crystal (index $\mu$) and each photon (index $j$) as:

\begin{align}\label{eqn5}
    \varphi_j^{(\mu)}(\omega_j)&=K^{(\mu)}_j(\omega_j)\ell_j^{(\mu)}+k_j(\omega_j)L\nonumber\\
    &=\frac{n_j^{(\mu)}(\omega_j)\,\ell_j^{(\mu)}\,\omega_j}{c}+k_j(\omega_j)L,
\end{align}

\noindent where $k_j(\omega_j)$ is the wavenumber associated with the dispersion from the nonlinear crystals themselves which does not have an index $\mu$ due to the fact that the crystals are assumed to be identical. $K_j^{(\mu)}$ is the wavenumber associated with the respective time delay, where the time delay is applied by a medium of lengths $\ell_j^{(\mu)}$ and refractive indices $n_j^{(\mu)}(\omega)$. The meaning of this term is that a photon with frequency $\omega_j$ passing through the material of the given length and refractive index will have its amplitude modulated by the factor $e^{i\varphi_j^{(\mu)}(\omega_j)}$. In the case of the NLI, each photon will then accumulate these factors depending upon which crystal the daughter photon pair was produced within. In particular, the modulation factor $\beta^{(\mu)}\equiv\prod_{j}\beta_j^{(\mu)}(\omega_j)$ modulates the joint spectrum according to the factors $e^{i\varphi_p^{(\mu)}(\omega_s+\omega_i)}$ which the pump acquired ``upstream''  of (prior to passing through) crystal $\mu$ and factors $e^{i\varphi_{s,i}^{(\mu)}(\omega_{s,i})}$ which signal and idler photons acquired ``downstream''  of (following passage through) crystal $\mu$. This can be written as:

\begin{align}\label{summation_rule}
        \beta^{(\mu)}&=\prod_{\text{upstream}} \exp\left({i\varphi_p^{(\text{upstream of }\mu)}(\omega_s+\omega_i)}\right)\,\nonumber\\
        &\qquad\times\prod_{\text{downstream}} \exp\left({i\varphi_s^{(\text{downstream of }\mu)}(\omega_s)}\right)\,\exp\left({i\varphi_i^{(\text{downstream of }\mu)}(\omega_i)}\right)\nonumber\\
        &=\prod_{m=1}^{\mu}\exp\left({i\varphi_p^{(m)}(\omega_s+\omega_i)}\right)\,\prod_{n=\mu}^N\exp\left({i\varphi_s^{(n)}(\omega_s)}\right)\,\exp\left({i\varphi_i^{(n)}(\omega_i)}\right)\nonumber\\
        &=\exp\left({i\sum_{m=1}^{\mu}\varphi_p^{(m)}(\omega_s+\omega_i)}\right)\,\exp\left({i\sum_{n=\mu}^N\varphi_s^{(n)}(\omega_s)}\right)\,\exp\left({i\sum_{n=\mu}^N\varphi_i^{(n)}(\omega_i)}\right).
\end{align}

We can expand this in terms of Eq. \eqref{eqn5}:

\begin{equation}\label{eqn8}
        \beta^{(\mu)}=\exp\left({i(N-1)(\Delta k+k_p)L}\right)\,\exp\left({-i(\mu-1)\Delta k L}\right)\,\beta_0^{(\mu)},
\end{equation}

\noindent where:

\begin{align}
    \beta_0^{(\mu)}=\exp\left(i\sum_{m=1}^{\mu}K_p^{(m)}\ell_p^{(m)}\right)\exp\left(i\sum_{n=\mu}^{N}K_s^{(n)}\ell_s^{(n)}\right)\exp\left(i\sum_{n=\mu}^{N}K_i^{(n)}\ell_i^{(n)}\right)
\end{align}

\noindent is the modulation term specifically due to the time delays. The full derivation of \eqref{eqn8} can be found in the appendix. The global phase $e^{i(N-1)(\Delta k + k_p)L}$ is ultimately unmeasurable and thus irrelevant. The factor $e^{-i(\mu-1)\Delta k L}$ would tend to affect the NLI interference in non-trivial ways, however, if the phase matching is sufficiently broad, we may approximate $\Delta k\approx 0$ over the spectral region of interest. For these reasons, we may neglect the parts of the phase factors which arise due to the nonlinear crystals and only consider the factors which are acquired by the time delay elements between the crystals, i.e. we may set $\beta^{(\mu)}=\beta_0^{(\mu)}$ for the purposes of all further discussion.

Finally, note that we can expand the phase $\varphi_{j}^{(\mu)}=K_j^{(\mu)}\ell_j^{(\mu)}$ as follows:

\begin{equation}
        K_j^{(\mu)}\ell_j^{(\mu)}=\omega_j\,\frac{n_j^{(\mu)}(\omega_j)\ell_j^{(\mu)}}{c}.
\end{equation}

We assume that the dispersion is negligible such that we can simplify this to obtain the phase factor in terms of a time delay $\tau_j^{(\mu)}$:

\begin{equation}
K_j^{(\mu)}\ell_j^{(\mu)}\approx\omega_j\,\frac{n_j^{(\mu)}(\omega_{j_0})\ell_j^{(\mu)}}{c}\equiv\omega_j\tau_j^{(\mu)}
.\end{equation} 

For simplicity, we will express all $\tau_j^{(\mu)}$ as multiples of a fundamental time delay denoted $\tau$ whose numerical value will be given where it is relevant.

\section{Nonlinear Interferometer Scheme}

The NLI scheme we develop consists of a series of four nonlinear crystals where we suppose that optics for applying separate delays between pump, signal, and idler are placed between the crystals. By simply choosing the sequence of time delays, either a grid state or a high-dimensional entangled state can be generated. We choose $N=4$ crystals as the minimum number of crystals that can produce both targeted states. While more crystals can define the spectral structures more clearly, they also add susceptibility to loss and dispersion, leading to diminishing returns and added complexity.

\subsection{Scheme for Generating Grid State}

\begin{figure}[!h]
\centering
    \includesvg[width=0.8\textwidth]{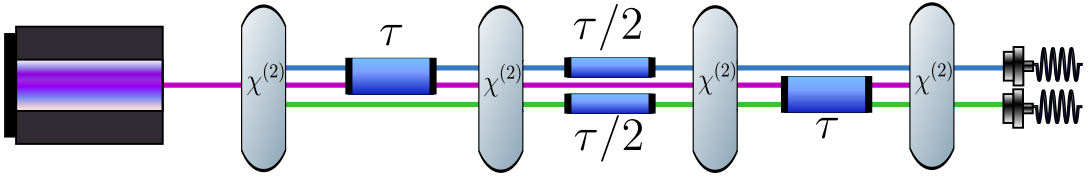}
    \caption{Block diagram of the nonlinear interferometer apparatus for generating the grid state. The fundamental time delay is $\tau=8.3\,\mathrm{ps}$. This is only a simplified version of the apparatus, not depicting actual optical components in the laboratory. Note that the pump (violet beam) does transmit through the final crystal, but we show it terminating there as a simplistic illustration of the fact that we filter it out at the end to leave only signal and idler beams at the final fiber couplings.}
    \label{fig:setup:grid}
\end{figure}

First, we treat the case for generating a grid state, with a delay sequence visualized in Fig. \ref{fig:setup:grid}. The sequence of time delays corresponding to a grid state modulation $\beta_{\text{grid}}(\omega_s,\omega_i)$ is:

\begin{align}\label{grid_state_time_delays}
        \{\tau_p^{(1)},\tau_p^{(2)},\tau_p^{(3)},\tau_p^{(4)}\}^{(\text{grid})}&=\{0,\tau,0,\tau\}\nonumber\\
        \{\tau_s^{(1)},\tau_s^{(2)},\tau_s^{(3)},\tau_s^{(4)}\}^{(\text{grid})}&=\{\tau,\frac{\tau}{2},0,0\}\nonumber\\
        \{\tau_i^{(1)},\tau_i^{(2)},\tau_i^{(3)},\tau_i^{(4)}\}^{(\text{grid})}&=\{0,\frac{\tau}{2},\tau,0\}.
\end{align}

$\tau_p^{(1)}$ and $\tau_{s,i}^{(4)}$ can only contribute a global phase to the overall interference term. The index $\mu$ relates to the fields involved in the PDC process in each the $\mu$th crystal, with the pump getting delayed prior to entering that crystal, and the signal/idler suffering delay after exiting that crystal. Applying the rule of Eq. \eqref{summation_rule} to this sequence of time delays to obtain $\{\beta^{(\mu)}_{\text{grid}}\}$ (recalling that $\varphi_j^{(\mu)}=\omega_j\tau_j^{(\mu)}$) and then summing over $\mu$ yields:

\begin{equation}\label{grid_modulation}
    \boxed{\beta_{\text{grid}}(\omega_s,\omega_i)=\sum_{\mu=1}^4\beta^{(\mu)}_{\text{grid}}=4\,e^{i\frac{3\omega_s\tau}{2}}\,e^{i2\omega_i\tau}\,\cos\frac{\left(\omega_s+\omega_i\right)\tau}{4}\,\cos\frac{\left(\omega_s-\omega_i\right)\tau}{4}}
\quad.\end{equation}

The combination of a modulation along the sum-frequency axis and another along the difference frequency axis causes the places where these modulation functions jointly have maxima to form a lattice structure with lattice spacing $d_{\text{grid}}={8\sqrt{2}\pi}/{\tau}$, such that modulation of an arbitrary joint spectral amplitude distribution by $\beta_{\text{grid}}$ forms the desired ``grid state''. Simulation of this modulation is shown in Fig. \ref{fig:fig3}.

The grid state can be expanded as:

\begin{equation}
|JSA_{\text{grid}}\rangle=\sum_{k,\ell}\mathcal{A}_{k,\ell}\,|\omega_{s}^{(k)},\omega_{i}^{(\ell)}\rangle
\quad ,\end{equation}

\noindent where the spectral state $|\omega_{j}^{(k)}\rangle$ ($j\in\{s,i\}$) denotes a state consisting of a single-peaked spectral function denoted $\psi$  centered at the single frequency $\omega_{{j}}^{(k)}\equiv\omega_0+kd_{\text{grid}}$, i.e.:

\begin{equation}
|\omega_{j}^{(k)}\rangle\equiv\int\mathrm{d}\omega_j\,\psi(\omega_j-\omega_{j}^{(k)})\,\hat{a}^{\dagger}_{\omega_j}|\text{vac}\rangle
.\end{equation}

%In general we suppose that:}

%\begin{align}
%    \psi(\omega)\approx\exp\left(-\frac{\omega^2}{4s^2}\right),
%\end{align}}

%for some bandwidth $s<\delta$.}
By making a frequency-resolved measurement of one photon of the down-converted pair after generating such a grid state, the total state can be projected onto a single-photon spectral superposition state. For instance, consider that we detect the signal photon and measure it to have frequency $\omega_{s}^{(0)}$, this projects the remaining idler photon onto the state:

\begin{equation}        |\Psi_i(\omega_i)\rangle=|\omega_{s}^{(0)}\rangle\langle\omega_{s}^{(0)}|JSA_{\text{grid}}\rangle=|\omega_{s}^{(0)}\rangle\otimes\sum_{\ell}\mathcal{B}_{\ell}|\omega_{i}^{(\ell)}\rangle,
\end{equation}

\noindent where $\mathcal{B}_{\ell}=\mathcal{A}_{0,\ell}$. This represents a high-dimensional spectral superposition state. This superposition of the basis states $|\omega_{i}^{(\ell)}\rangle$ could be used in quantum computing and quantum key distribution protocols. 

\subsection{Scheme for Generating High-Dimensional Entangled State}

\begin{figure}[!h]
\centering
    \includesvg[width=0.8\textwidth]{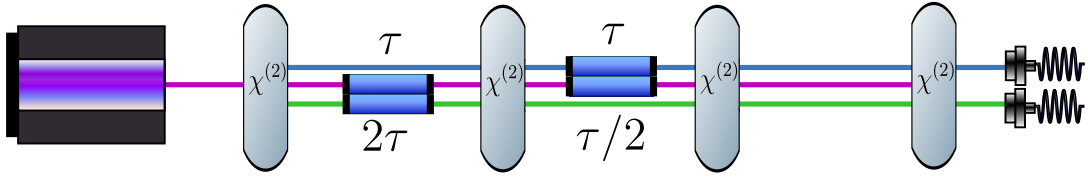}
    \caption{Block diagram of the nonlinear interferometer apparatus for generating the high-dimensional entangled state. The fundamental time delay is $\tau=1.0\,\mathrm{ps}$.}
    \label{fig:setup:hde}
\end{figure}

In a similar fashion, the sequence of time delays that gives rise to an HDE state as shown in Fig. \ref{fig:setup:hde} is:

\begin{align}\label{HDE_state_time_delays}
        \{\tau_p^{(1)},\tau_p^{(2)},\tau_p^{(3)},\tau_p^{(4)}\}^{(\text{HDE})}&=\{0,\tau,\frac{\tau}{2},0\}\nonumber\\
        \{\tau_s^{(1)},\tau_s^{(2)},\tau_s^{(3)},\tau_s^{(4)}\}^{(\text{HDE})}&=\{0,\tau,0,0\}\nonumber\\
        \{\tau_i^{(1)},\tau_i^{(2)},\tau_i^{(3)},\tau_i^{(4)}\}^{(\text{HDE})}&=\{2\tau,0,0,0\}.
\end{align}

There is no additional delay applied to any of the fields between third and fourth crystals in this scheme, however this does not mean that the fourth crystal is optional. With the fourth crystal in place with delays $\tau_p^{(4)}=\tau_{s,i}^{(3)}=0$, which is not equivalent to $\beta^{(4)}=0$ by the definition of $\beta^{(\mu)}$ from \eqref{summation_rule}; the fourth crystal certainly affects the overall interference term, which suppresses phasematched spectral components outside the main islands, which may not be desirable for the HDE state. Once again by applying the rule of \eqref{summation_rule}, we obtain that:

\begin{equation}\label{HDE_modulation}
    \boxed{\beta_{\text{HDE}}(\omega_s,\omega_i)=\sum_{\mu=1}^4\beta_{\text{HDE}}^{(\mu)}=4\,e^{i\frac{3(\omega_s+\omega_i)\tau}{2}}\,\cos^2\frac{\left(\omega_s-\omega_i\right)\tau}{4}}
.\end{equation}

The difference frequency term $\omega_s-\omega_i$ causes modulation along the positive diagonal of a joint spectral plot. When a joint spectrum that lies nearly along the anti-diagonal is modulated by a function whose contours lie along this axis, the joint spectral intensity is constrained to a series of joint spectral ``islands'' along the anti-diagonal. When written as a quantum state, this is a high-dimensional Bell state where the basis states are approximately Gaussian spectral modes:

\begin{equation}\label{HDE state}
    |JSA_{\text{HDE}}\rangle=\sum_k\mathcal{C}_k|\omega_{s}^{(k)},\omega_{i}^{(-k)}\rangle,
\end{equation}

\noindent where $\omega_{{s,i}}^{(k)}=\omega_0+kd_{\text{HDE}}$ and $d_{\text{HDE}}\approx{4\pi}/{\sqrt{2}\tau}$ (the equality is exact if the unmodulated JSI has its maximum intensity along the $-45^{\circ}$ sum-frequency axis). The state given by Eq. \eqref{HDE state} is then a high-dimensional entangled state. Simulations of the HDE state generation are shown in Fig. \ref{fig:fig4}.
A detailed derivation of the modulation functions given in Eq. \eqref{grid_modulation} and \eqref{HDE_modulation} is given in the supplemental document.

%%%%%%%%%%%%%%%%%%%%%%%%%%%%%%%%%%%%%%%%%%%%%%%%%%%%%%%%
\section{Simulation Results}

In the simulations the phase matching is assumed to be sufficiently broad such that the condition $\varphi_j^{(\mu)}= K_j^{(\mu)}\ell_j^{(\mu)}=\omega_j\tau_j^{(\mu)}$ is met, but we neglected dispersion arising from the nonlinear crystals themselves and only calculated the modulation based on the time delays. In our simulations, we assumed a Gaussian pump spectrum with $\lambda_{p_0}=775\,\mathrm{nm}$ and $\sigma_p=2.0\,\mathrm{nm} $ (FWHM=4.71 nm).

The simulation works by first calculating $JSA_0$ assuming a Gaussian pump spectrum  and a sinc-shaped phasematching function as well as perfect energy conservation $(\omega_p=\omega_s+\omega_i)$. To calculate the modulation functions, the rule of \eqref{summation_rule} is included explicitly, along with sets of time delays. The program then calculates and adds the phase factors $\beta_j^{(\mu)}$ in the appropriate manner to obtain the modulation term -- in other words, the modulation functions $\beta_{\text{grid, HDE}}$ are directly simulated rather than being programmed-in wholesale. When we include the effects of crystal dispersion, we instead do so by directly imposing the result of Eq. \eqref{eqn8} on each factor $\beta^{(\mu)}$ directly, still neglecting the global phase part of that expression.

\begin{figure}
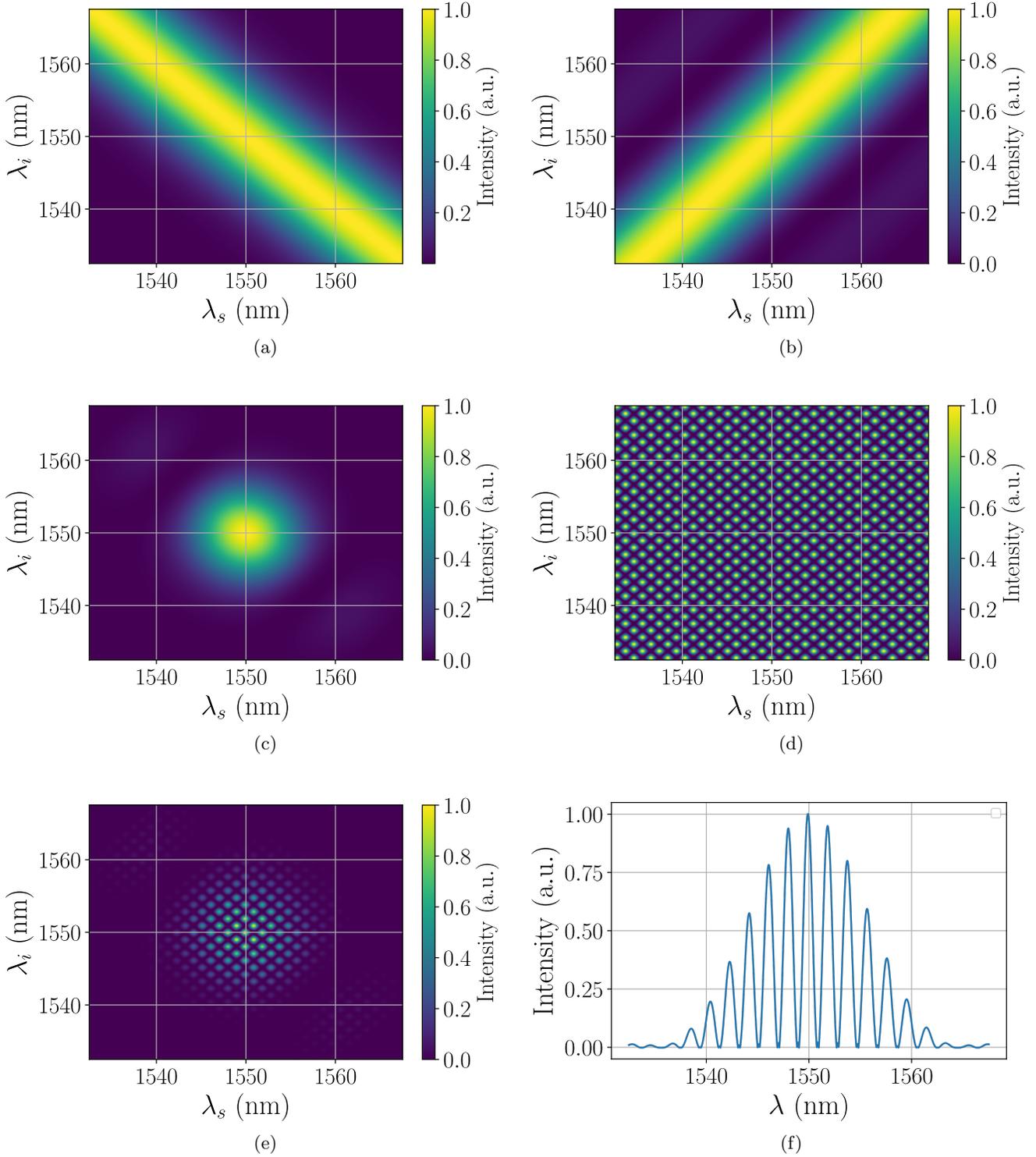

    \centering 
    \subfloat[]{\includesvg[width=0.5\textwidth]{PEF_grid.svg}}
    \subfloat[]{\includesvg[width=0.5\textwidth]{PMF_grid.svg}}\\
    \subfloat[]{\includesvg[width=0.5\textwidth]{JSI_0_grid.svg}}
    \subfloat[]{\includesvg[width=0.5\textwidth]{F_mod_grid.svg}}\\
    \subfloat[]{\includesvg[width=0.5\textwidth]{grid_0_dB.svg}}
    \subfloat[]{\includesvg[width=0.5\textwidth]{Idler_spectrum.svg}}
    \caption{The joint spectral grid states along with the pump, phase matching, and modulation functions and the projected idler photon state. (a) The norm-square of the pump envelope function for the grid state. (b) The norm-square of the phase matching function for the grid state. (c) The unmodulated joint spectral intensity for the grid state. (d) The norm square of the modulation function for the grid state displayed with pseudo-normalization. (e) The grid state . (f) The norm-square of the grid state where projective measurement projects onto $\lambda_s=1550\,\text{nm}$.}
    \label{fig:fig3}
\end{figure}

To simulate a grid state, we use type-II ($x\rightarrow xz$) SPDC taking place within a periodically-poled Potassium Titanyl Phosphate (ppKTP) crystal of length $L=1.0\,\mathrm{mm}$, using Sellmeier equations as found in \cite{KTP_Sellmeier} to obtain expressions for $k_j(\omega_j)$ and with poling period ($\Lambda$) chosen to quasi-phase match the degenerate downconversion 775 nm $\rightarrow$ (1550 nm, 1550 nm) and with time delay $\tau=8.3\,\mathrm{ps}$. In Fig. \ref{fig:fig3} we show results for phasematching function (\ref{fig:fig3}a), pump envelope function (Fig. \ref{fig:fig3}b), unmodulated JSI (Fig. \ref{fig:fig3}c), modulation function (Fig. \ref{fig:fig3}d), the full JSI of the nonlinear interferometer (Fig. \ref{fig:fig3}e), and finally the spectrum of the projection in Fig. \ref{fig:fig3}f.

\begin{figure}
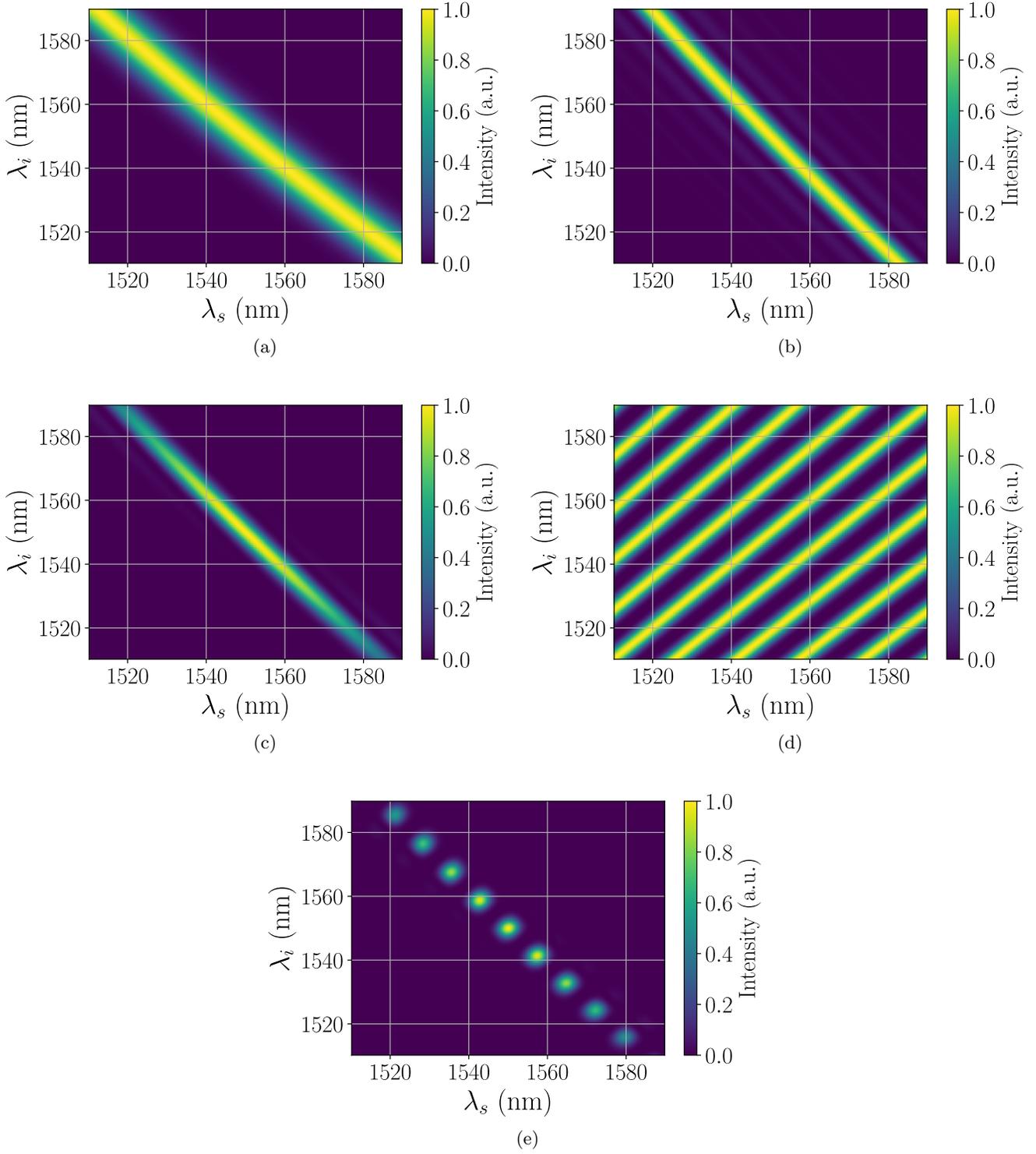

    \centering 
    \subfloat[]{\includesvg[width=0.5\textwidth]{PEF_HDE.svg}}
    \subfloat[]{\includesvg[width=0.5\textwidth]{PMF_HDE.svg}}\\
    \subfloat[]{\includesvg[width=0.5\textwidth]{JSI_0_HDE.svg}}
    \subfloat[]{\includesvg[width=0.5\textwidth]{F_mod_HDE.svg}}\\
    \subfloat[]{\includesvg[width=0.5\textwidth]{HDE_0_dB.svg}}
    \caption{The HDE joint spectrum. (a) Norm square of the pump envelope function for the HDE state. (b) Norm square of the phase matching function for the HDE state. (c) The unmodulated joint spectral intensity for the HDE state. (d) Modulation function for the HDE state. (e) The modulated joint spectral intensity HDE state.}
    \label{fig:fig4}
\end{figure}

In the case of the HDE state simulations, we employ SPDC within a periodically-poled Lithium Niobate (ppLN) crystal of length $L=1.0\,\text{mm}$  and type-II (eoe) phase matching with critical phase matching such that \cite{APM}:

\begin{equation}
    n_e(\lambda,\theta)=\left(\frac{\cos^2\theta}{n_o^2(\lambda)}+\frac{\sin^2\theta}{n_{e}^2(\lambda,\,90^{\circ})}\right)^{-1/2},
\end{equation}

\noindent with phase matching angle $\theta=30^{\circ}$.  The fundamental time delay used in the case of generating the HDE state is $\tau=1.0\,\mathrm{ps}$. In Fig. \ref{fig:fig4} we show results for phasematching function (\ref{fig:fig4}a), pump envelope function (Fig. \ref{fig:fig4}b), unmodulated JSI (Fig. \ref{fig:fig4}c), modulation function (Fig. \ref{fig:fig4}d), and finally the full JSI of the nonlinear interferometer (Fig. \ref{fig:fig4}e).

%%%%%%%%%%%%%%%%%%%%%%%%%%%%%%%%%%%%%%%%%%%%%%%%%%%%%%%%%%%%%%%
\section{Analysis of Loss}

Any implementation of the scheme presented here would suffer some degree of loss or lack of mode overlap between crystals. Mode overlap could for example be affected by waveguide mode mismatch or walk-ff within the crystals. Waveguide losses, inefficient optical fiber coupling, reflections at optical interfaces, and mode matching could all affect visibility in the NLI. Practically, loss of overlap and linear loss both result in poor visibility, which is why we only model loss in an explicit way. An analysis of loss within NLI's in the context of their application to quantum sensing has also been performed by \cite{NLI_seeding}. The effect of linear loss can also be seen as equivalent to the effect of spatial mode distortion, such that the addition of loss to the simulations may be viewed as equivalent to the addition of spatial mode distortion.

Thus, for a loss ($X$) expressed in dB at every optical interface, the phase factor $\beta^{(\mu)}$ will be modulated such that the crystal-specific phase factor $\beta^{(\mu)}$ will become $(10^{-\frac{X}{20}})^{(2\mu-1)}\,\beta^{(\mu)}$.

We simulate the effect of this loss for an interval of $X=[0\,\text{dB},\,20\,\text{dB}]$ and use the Schmidt number as a metric as well as the intensity overlap defined as:

\begin{equation}
    \mathcal{O}(s_a,s_b)=\int\mathrm{d}\lambda_1\,\mathrm{d}\lambda_2\,|s_a(\lambda_1,\lambda_2)|^2\,|s_b(\lambda_1,\lambda_2)|^2\quad
,\end{equation}

\begin{figure}
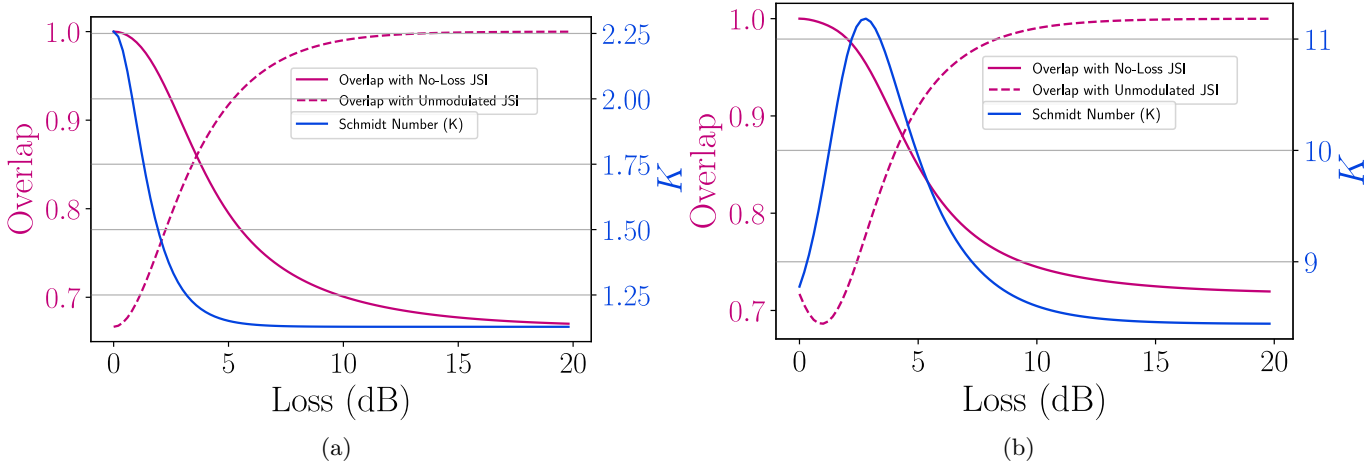

    \centering
    \subfloat[]{
    \includesvg[width=0.5\textwidth]{Overlap_and_K_vs_Loss_grid}}
    \subfloat[]{
    \includesvg[width=0.5\textwidth]{Overlap_and_K_vs_Loss_HDE}}
\caption{The results of numerical analysis of the behavior of the Schmidt number as a function of loss (blue line), overlap of the lossy state with the lossless state vs. loss (solid magenta line), and overlap of the lossy state with the unmodulated state (dashed magenta lines) for the grid state (a) and HDE state (b).}
\label{fig:overlap}
\end{figure}

\noindent where $\{s_a,s_b\}$ are normalized joint spectral states. The results are shown in Fig. \ref{fig:overlap}a for the grid state, and in Fig. \ref{fig:overlap}b for the HDE state. We choose to evaluate this overlap between two pairs of joint spectral states: the overlap between the unmodulated joint spectrum $JSI_0$ and the spectrum with variable loss, and the overlap between the spectrum with no loss and the spectrum with variable loss (dashed and solid magenta lines in Fig. \ref{fig:overlap}, respectively). 
As the amount of loss per optical interface is increased, the interference visibility is diminished and thus the overlap between the unmodulated $JSI_0$ and the NLI-modulated state approaches unity while the overlap between the lossy modulated JSI and the lossless JSI diminishes. In the case of the grid state, the unmodulated JSI is very nearly a spectrally pure state with $K\approx 1$. The loss drives the state toward that single-mode state and the Schmidt number thus decreases toward 1 with increasing loss. In the case of the HDE state, the unmodulated state is multimode. As loss increases, the Schmidt number first increases, then eventually settles toward the Schmidt number associated with the unmodulated multimode state.

We can see this behavior qualitatively in the JSIs under loss shown in Fig. \ref{fig:fig6}a-c, where the space between the features of the grid state gradually fills in as loss is increased from 0 dB to 10 dB, approaching the unmodulated JSI. To be useful for spectral multiplexing, the HDE state should be separable in intensity. The loss causes the individual islands to begin to blur together, approaching the unmodulated, zero-visibility case in the limit as loss goes toward infinity. The projection of the grid state via projective measurement on the signal photon shown in seen in Fig. \ref{fig:fig6}(d-f) in comparison to the lossless state is somewhat robust against loss, although the fringe visibility decreases as loss is increased. For the HDE state's JSI under loss shown in Fig. \ref{fig:fig6}g-i, we can also observe that the JSI fills in and becomes more similar to the unmodulated JSI. Islands are not clearly distinguishable in intensity.

\begin{figure}
    \centering
    \subfloat[]{
    \includesvg[width=0.33\textwidth]{grid_0_dB}}
    \subfloat[]{
    \includesvg[width=0.33\textwidth]{grid_4_dB}}
    \subfloat[]{
    \includesvg[width=0.33\textwidth]{grid_10_dB}}\\
    \subfloat[]{
    \includesvg[width=0.33\textwidth]{proj_0_dB}}
    \subfloat[]{
    \includesvg[width=0.33\textwidth]{proj_4_dB}}
    \subfloat[]{
    \includesvg[width=0.33\textwidth]{proj_10_dB}}\\
    \subfloat[]{
    \includesvg[width=0.33\textwidth]{HDE_0_dB}}
    \subfloat[]{
    \includesvg[width=0.33\textwidth]{HDE_4_dB}}
    \subfloat[]{
    \includesvg[width=0.33\textwidth]{HDE_10_dB}}
\caption{A selection of joint spectral intensities and projected idler states for grid and HDE for varying amounts of loss. (a)-(c): grid states for 0.0, 4.0, and 10.0 dB of loss, respectively. (d)-(f): the state of the idler photon obtained via projective measurement and postselection of $\lambda_s=1550\,\text{nm}$ for 0.0, 4.0, and 10.0 dB of loss. (g)-(i): The HDE state for 0.0, 4.0, and 10.0 dB of loss.}
\label{fig:fig6}
\end{figure}

The behavior of the joint spectral modes of the grid state as a function of loss is interesting, and is shown in Fig. \ref{fig:fig7}. The left column (panels a, c, e) show the first mode for each photon for loss of 0, 1 and 3 dB, and the right column (panels b, d, f) show the second modes. Without loss, interference fringes are visible in the first two modes of both photons, which represent $\sim99\%$ of the total mode decomposition. However, with even modest amounts of loss, the fringes remain visible in the first mode of one photon, while the other mode becomes a Gaussian spectrum, then in the next mode this behavior is exactly reversed. It can be seen that the qudit spectral structure that is desired is destroyed even for just 1 dB of loss, with an entirely different modal structure emerging.

\begin{figure}
    \centering
    \subfloat[]{
    \includesvg[width=0.5\textwidth]{grid_SVD_mode_0_0.0_dB_loss}}
    \subfloat[]{
    \includesvg[width=0.5\textwidth]{grid_SVD_mode_1_0.0_dB_loss}}\\
    \subfloat[]{
    \includesvg[width=0.5\textwidth]{grid_SVD_mode_0_1.0_dB_loss}}
    \subfloat[]{
    \includesvg[width=0.5\textwidth]{grid_SVD_mode_1_1.0_dB_loss}}\\
    \subfloat[]{
    \includesvg[width=0.5\textwidth]{grid_SVD_mode_0_3.0_dB_loss}}
    \subfloat[]{
    \includesvg[width=0.5\textwidth]{grid_SVD_mode_1_3.0_dB_loss}}
\caption{The first two signal and idler modes calculated via singular value decomposition for varying amount of loss. (a) Signal and idler mode 0 of the grid state in the absence of loss. (b) Signal and idler mode 1 of the grid state in the absence of loss. (c) Signal and idler mode 0 of the grid state with 1.0 dB of loss. (d) Signal and idler mode 1 of the grid state with 1.0 dB of loss. (e) Signal and idler mode 0 of the grid state with 3.0 dB of loss. (f) Signal and idler mode 1 of the grid state with 3.0 dB of loss.}
\label{fig:fig7}
\end{figure}

%%%%%%%%%%%%%%%%%%%%%%%%%%%%%%%%%%%%%%%%%%%%%%%%%%%%%%%%%%%%%%%%%%%%%%%%%
\section{Analysis of Crystal Dispersion}

\begin{figure}[!h]
    \centering 
    \subfloat[]{\includesvg[width=0.5\textwidth]{grid_0_dB_with_crystal_dispersion.svg}}
    \subfloat[]{\includesvg[width=0.5\textwidth]{HDE_with_disp_LN.svg}}\\
    \subfloat[]{\includesvg[width=0.5\textwidth]{grid_with_disp_LN.svg}}
    \subfloat[]{\includesvg[width=0.5\textwidth]{proj_with_disp_LN.svg}}
    \caption{{Simulations of the grid and HDE modulation functions and joint spectral intensities with zero loss, but including crystal dispersion. (a) The grid state JSI with crystal dispersion,  (b) the HDE with crystal dispersion, (c) grid state with different choice of pump and phasematching that masks off-island intensity, (d) projected spectrum for the modified grid state.}}
    \label{fig:fig8}
\end{figure}

To investigate the effect of dispersion in the model, we perform the simulations from the zero-loss case but without the assumption of $\Delta k=0$ over the entire spectrum, so that crystal dispersion in Eq. \eqref{eqn8} becomes relevant. The results are shown in Fig. \ref{fig:fig8}. In panel a, we show the JSI of the grid state under crystal dispersion. Features on the edges of the JSI are filling in as a result. For the HDE state shown in panel b, we can always choose a crystal length short enough for dispersion to have no effect, with the pump and modulation shaping the JSI, decoupled from crystal length. A largely unchanged JSI compared to the case without dispersion is achieved. We decreased the pump bandwidth from $\sigma_p=1\,\text{nm}$ to $\sigma_p=0.25\,\text{nm}$ and the ppLN crystal length from $L=1\,\text{nm}$ to $L=0.5\,\text{nm}$ and looked over a narrower wavelength range when plotting. We also increased the fundamental time delay in the case of the HDE state from $\tau=1.0\,\text{ps}$ to $\tau\approx 2.94\,\text{ps}$.

We attempt to mitigate the effect of crystal dispersion in a similar way for the grid state. pump bandwidth and phasematching bandwidth are coupled to achieve the underlying unmodulated JSI. The heralded spectrum can be salvaged by choosing the same crystal and pump parameters as in the improved HDE simulations (ppLN crystal,  $\sigma_p=0.25\,\text{nm}$, $L=0.5\,\text{nm}$, $\theta=30\degree$, $\tau\approx 40.0\,\text{ps}$). The most significant change in this case is that the form of $JSI_0$ with these parameters is no longer a single-mode island, but instead the highly-correlated, multimode joint spectrum already shown in Fig. \ref{fig:fig8}c. When the unmodulated joint spectrum acquires the elongated, multimode form, the result of a projection along one column of the grid state in fact changes very little. An almost ideal heralded spectrum merges as shows in Fig. \ref{fig:fig8}d. Compared to the dispersion-less case, the herald measurement needs to spectrally resolve one row of the JSI, introducing loss.

\color{black}\section{Discussion and Conclusion}

We demonstrate that a nonlinear interferometer consisting of four nonlinear crystals combined with linear optical elements to impart linear spectral phases can generate two useful classes of photon pair states in the spectral basis, namely high-dimensional entangled states and grid states. Frequency-resolving postselection upon one photon of the the grid state yields a multi-peaked spectral state on its sibling photon. This state can be approximated well by a superposition of Gaussian spectra of different central frequencies, where these central frequencies are evenly spaced. These spectral features can be regarded as spectral qudit states, which are desirable for their applicability to quantum computing due to their high-dimensional nature. 
The same scheme can generate high-dimensional entangled states by modulating a joint spectrum that lies along the anti-diagonal axis with a cosine with contours along the difference frequency axis. Such high-dimensional entangled states have applicability in quantum communications. In particular, they could be used to achieve high-dimensional quantum teleportation schemes. Such states can also be used for signal multiplexing, which is useful in achieving a higher photon bit rate through down-conversion schemes while keeping the probability amplitudes for higher-order, multi-pair terms in the expansion negligible. Multiplexing overcomes a fundamental tradeoff between the bit rate achievable in schemes utilizing pair generation processes and the photon number purity of these processes  \cite{spectral_multiplexing_5,spectral_multiplexing_3,spectral_multiplexing_4,spectral_multiplexing_6,power_number_fidelity_threshold_and_SMUX,multi-pair-statistics,spectral_multiplexing_1,spectral_multiplexing_2}. Spectrally multiplexed states also have application to quantum communications schemes utilizing wavelength division multiplexing \cite{WDM1,WDM2}. Moreover, the single photon spectral mode superposition states may be useful in certain experiments probing foundational questions \cite{doi:10.1073/pnas.1921529118,apr_2}, although the effects of dispersion and loss on the grid state are likely problematic for fashioning quantum-analogue experiments to probe superoscillatory behavior.

We show how the ability to generate these states changes when loss at optical interfaces is introduced into the simulations. The joint spectral states for both grid and HDE remain qualitatively similar to the zero-loss case for modest amounts of loss, however the modes and -- in the case of the grid state -- the projected idler spectrum change non-trivially with the loss. 
With modest amounts of loss, the projected idler spectrum acquires ancillary peaks that will quickly interfere with their utility as spectral superposition states in quantum computing and other areas. Future work may be needed to show how this scheme can be made more robust against loss, perhaps by compensatory modulation of the pump spectrum with an optical pulse shaper. On the other hand, the HDE states remain relatively robust against moderate amounts of loss and may remain useful, although the overlap of the islands may quickly contaminate any usefulness in spectral multiplexing, as the islands cannot be separated in intensity alone. Possibly, a combination of schemes involving both shaping of the pump spectrum and crystal domain engineering \cite{Shukhin:24} along with nonlinear interferometry could provide ultimate control over the biphoton spectrum and allow for many degrees of freedom.

Implementation of a four-stage nonlinear interferometer requires exceptional interferometric stability, which has been demonstrated in a number of ways. Su et. al have implemented a three-stage interferometer using fiber components \cite{Li:19}. In recent work on a source for polarization-entangled photon pairs, Grayson et al. combined a free-space interferometer with four spatial modes on a monolithic mount \cite{grayson2026highperformancesourceindistinguishablepolarizationentangled}. Integrated quantum optics platforms such as thin-film Lithium Niobate \cite{Zhu:21} offer the stability of monolithic integration, low loss, and guaranteed mode overlap.

\section*{Funding}
This work is funded by the National Science Foundation under Grant No. 2427047

\section*{Disclosures}
The authors declare no conflicts of interest.

\section*{Data availability} All data generated are part of the manuscript. Python code used for the simulations is available at \url{https://doi.org/10.5281/zenodo.19371079}.

\section*{Bibliography}
\bibliography{citations}

\appendix
\section{Derivation of the Interference Terms}

Here we show a detailed derivation of how the modulation functions arise given the stated sequences of time delays.

Consider again the sequence of time delays for the grid state:

\begin{align}
        \{\tau_p^{(1)},\tau_p^{(2)},\tau_p^{(3)},\tau_p^{(4)}\}^{(\text{grid})}&=\{0,\tau,0,\tau\}\nonumber\\
        \{\tau_s^{(1)},\tau_s^{(2)},\tau_s^{(3)},\tau_s^{(4)}\}^{(\text{grid})}&=\{\tau,\frac{\tau}{2},0,0\}\nonumber\\
        \{\tau_i^{(1)},\tau_i^{(2)},\tau_i^{(3)},\tau_i^{(4)}\}^{(\text{grid})}&=\{0,\frac{\tau}{2},\tau,0\}.
\end{align}

Applying the rule of Eq. 13 we obtain:

\begin{align}
\beta_{\text{grid}}^{(1)}&=e^{i\omega_p\tau_p^{(1)}}\,e^{i\omega_s(\tau_s^{(1)}+\tau_s^{(2)}+\tau_s^{(3)}+\tau_s^{(4)})}\,e^{i\omega_i(\tau_i^{(1)}+\tau_i^{(2)}+\tau_i^{(3)}+\tau_i^{(4)})}\nonumber\\
&=e^{i(\omega_s+\omega_i)(0)}\,e^{i\omega_s(\tau+\frac{\tau}{2}+0+0)}\,e^{i\omega_i(0+\frac{\tau}{2}+\tau+0)}\nonumber\\
&=\boxed{e^{i\frac{3\omega_s\tau}{2}}\,e^{i\frac{3\omega_i\tau}{2}}}.
\end{align}

\begin{align}        \beta_{\text{grid}}^{(2)}&=e^{i\omega_p(\tau_p^{(1)}+\tau_p^{(2)})}\,e^{i\omega_s(\tau_s^{(2)}+\tau_s^{(3)}+\tau_s^{(4)})}\,e^{i\omega_i(\tau_i^{(2)}+\tau_i^{(3)}+\tau_i^{(4)})}\nonumber\\
        &=e^{i(\omega_s+\omega_i)(0+\tau)}\,e^{i\omega_s(\frac{\tau}{2}+0+0)}\,e^{i\omega_i(\frac{\tau}{2}+\tau+0)}\nonumber\\
        &=\boxed{e^{i\frac{3\omega_s\tau}{2}}\,e^{i\frac{5\omega_i\tau}{2}}}.
\end{align}

\begin{align}        \beta_{\text{grid}}^{(3)}&=e^{i\omega_p(\tau_p^{(1)}+\tau_p^{(2)}+\tau_p^{(3)})}\,e^{i\omega_s(\tau_s^{(3)}+\tau_s^{(4)})}\,e^{i\omega_i(\tau_i^{(3)}+\tau_i^{(4)})}\nonumber\\
        &=e^{i(\omega_s+\omega_i)(0+\tau+0)}\,e^{i\omega_s(0+0)}\,e^{i\omega_i(\tau+0)}\nonumber\\
        &=\boxed{e^{i{\omega_s\tau}}\,e^{i2\omega_i\tau}}.
\end{align}

\begin{align}        \beta_{\text{grid}}^{(4)}&=e^{i\omega_p(\tau_p^{(1)}+\tau_p^{(2)}+\tau_p^{(3)}+\tau_p^{(4)})}\,e^{i\omega_s(\tau_s^{(4)})}\,e^{i\omega_i(\tau_i^{(4)})}\nonumber\\
        &=e^{i(\omega_s+\omega_i)(0+\tau+0+\tau)}\,e^{i\omega_s(0)}\,e^{i\omega_i(0)}\nonumber\\
        &=\boxed{e^{i2{\omega_s\tau}}\,e^{i2\omega_i\tau}}.
\end{align}

Summing over all the crystals we thus obtain:

\begin{align}
        \beta_{\text{grid}}&\equiv\sum_{\mu=1}^{4}\beta_{\text{grid}}^{(\mu)}\nonumber\\
        &=e^{i\frac{3\omega_s\tau}{2}}\,e^{i\frac{3\omega_i\tau}{2}}+e^{i\frac{3\omega_s\tau}{2}}\,e^{i\frac{5\omega_i\tau}{2}}+e^{i{\omega_s\tau}}\,e^{i2\omega_i\tau}+e^{i2{\omega_s\tau}}\,e^{i2\omega_i\tau}\nonumber\\
        &=e^{i\frac{3\omega_s\tau}{2}}\,e^{i2\omega_i\tau}\,\left(\underbrace{e^{-i\frac{\omega_i\tau}{2}}+e^{i\frac{\omega_i\tau}{2}}}_{2\cos\frac{\omega_i\tau}{2}}+\underbrace{e^{-i\frac{\omega_s\tau}{2}}+e^{i\frac{\omega_s\tau}{2}}}_{2\cos\frac{\omega_s\tau}{2}}\right)\nonumber\\
        &=2\,e^{i\frac{3\omega_s\tau}{2}}\,e^{i2\omega_i\tau}\,\left(\cos\frac{\omega_s\tau}{2}+\cos\frac{\omega_i\tau}{2}\right)\nonumber\\
        &\overset{\text{sum to product}}{=}\boxed{4\,e^{i\frac{3\omega_s\tau}{2}}\,e^{i2\omega_i\tau}\,\cos\frac{(\omega_s+\omega_i)\tau}{4}\cos\frac{(\omega_s-\omega_i)\tau}{4}}.
\end{align}

Following the same procedure for $\beta_{\text{HDE}}$, we revisit the sequence of HDE time delays:

\begin{align}
        \{\tau_p^{(1)},\tau_p^{(2)},\tau_p^{(3)},\tau_p^{(4)}\}^{(\text{HDE})}&=\{0,\tau,\frac{\tau}{2},0\}\nonumber\\
        \{\tau_s^{(1)},\tau_s^{(2)},\tau_s^{(3)},\tau_s^{(4)}\}^{(\text{HDE})}&=\{0,\tau,0,0\}\nonumber\\
        \{\tau_i^{(1)},\tau_i^{(2)},\tau_i^{(3)},\tau_i^{(4)}\}^{(\text{HDE})}&=\{2\tau,0,0,0\}.
\end{align}

Therefore:

\begin{align}        \beta_{\text{HDE}}^{(1)}&=e^{i\omega_p\tau_p^{(1)}}\,e^{i\omega_s(\tau_s^{(1)}+\tau_s^{(2)}+\tau_s^{(3)}+\tau_s^{(4)})}\,e^{i\omega_i(\tau_i^{(1)}+\tau_i^{(2)}+\tau_i^{(3)}+\tau_i^{(4)})}\nonumber\\
        &=e^{i(\omega_s+\omega_i)(0)}\,e^{i\omega_s(0+\tau+0+0)}\,e^{i\omega_i(2\tau+0+0+0)}\nonumber\\
        &=\boxed{e^{i\omega_s\tau}\,e^{i2\omega_i\tau}}.
\end{align}

\begin{align}        \beta_{\text{HDE}}^{(2)}&=e^{i\omega_p(\tau_p^{(1)}+\tau_p^{(2)})}\,e^{i\omega_s(\tau_s^{(2)}+\tau_s^{(3)}+\tau_s^{(4)})}\,e^{i\omega_i(\tau_i^{(2)}+\tau_i^{(3)}+\tau_i^{(4)})}\nonumber\\
        &=e^{i(\omega_s+\omega_i)(0+\tau)}\,e^{i\omega_s(\tau+0+0)}\,e^{i\omega_i(0+0+0)}\nonumber\\
        &=\boxed{e^{i2\omega_s\tau}\,e^{i\omega_i\tau}}.
\end{align}

\begin{align}       \beta_{\text{HDE}}^{(3)}&=e^{i\omega_p(\tau_p^{(1)}+\tau_p^{(2)}+\tau_p^{(3)})}\,e^{i\omega_s(\tau_s^{(3)}+\tau_s^{(4)})}\,e^{i\omega_i(\tau_i^{(3)}+\tau_i^{(4)})}\nonumber\\
        &=e^{i(\omega_s+\omega_i)(0+\tau+\frac{\tau}{2})}\,e^{i\omega_s(0+0)}\,e^{i\omega_i(0+0)}\nonumber\\
        &=\boxed{e^{i\frac{3\omega_s\tau}{2}}\,e^{i\frac{3\omega_i\tau}{2}}}.
\end{align}

\begin{align}     \beta_{\text{HDE}}^{(4)}&=e^{i\omega_p(\tau_p^{(1)}+\tau_p^{(2)}+\tau_p^{(3)}+\tau_p^{(4)})}\,e^{i\omega_s(\tau_s^{(4)})}\,e^{i\omega_i(\tau_i^{(4)})}\nonumber\\
        &=e^{i(\omega_s+\omega_i)(0+\tau+\frac{\tau}{2}+0)}\,e^{i\omega_s(0)}\,e^{i\omega_i(0)}\nonumber\\
        &=\boxed{e^{i\frac{3\omega_s\tau}{2}}\,e^{i\frac{3\omega_i\tau}{2}}}.
\end{align}

Therefore:

\begin{align}
    \beta_{\text{HDE}}&=\sum_{\mu=1}^{4}\beta_{\text{HDE}}^{(\mu)}\nonumber\\
    &=e^{i\omega_s\tau}\,e^{i2\omega_i\tau}+e^{i2\omega_s\tau}\,e^{i\omega_i\tau}+e^{i\frac{3\omega_s\tau}{2}}\,e^{i\frac{3\omega_i\tau}{2}}+e^{i\frac{3\omega_s\tau}{2}}\,e^{i\frac{3\omega_i\tau}{2}}\nonumber\\
    &=e^{i\frac{3(\omega_s+\omega_i)\tau}{2}}\,\left(\underbrace{e^{-i\frac{(\omega_s-\omega_i)\tau}{2}}+e^{+i\frac{(\omega_s-\omega_i)\tau}{2}}}_{2\cos\frac{(\omega_s-\omega_i)\tau}{2}}+2\right)\nonumber\\
    &=2\,e^{i\frac{3(\omega_s+\omega_i)\tau}{2}}\,\left(1+\cos\frac{(\omega_s-\omega_i)\tau}{2}\right)\nonumber\\
    &=\boxed{4\,e^{i\frac{3(\omega_s+\omega_i)\tau}{2}}\,\cos^2\frac{(\omega_s-\omega_i)\tau}{4}}.
\end{align}

\section{Derivation of Equation (14)}

Here we show a the full detailed derivation of the expression of equation (14) in the main text.

\begin{align}\label{eqnS13}
        \beta^{(\mu)}&=\exp\left({i\sum_{m=1}^{\mu}(K_p^{(m)}\ell_p^{(m)}+\sum_{m=1}^{\mu-1}k_pL)}\right)\,\nonumber\\
        &\qquad\times\exp\left({i\sum_{n=\mu}^N(K_s^{(n)}\ell_s^{(n)}+\sum_{n=\mu+1}^Nk_sL)}\right)\,\exp\left({i\sum_{n=\mu}^N(K_i^{(n)}\ell_i^{(n)}+\sum_{n=\mu+1}^Nk_iL)}\right)\nonumber\\
        &=\left[\exp\left({i\sum_{m=1}^{\mu-1}k_pL}\right)\,\exp\left({i\sum_{n=\mu+1}^N(k_s+k_i)L}\right)\right]\,\nonumber\\
        &\qquad\times\underbrace{\exp\left({i\sum_{m=1}^{\mu}K_p^{(m)}\ell_p^{(m)}}\right)\,\exp\left({i\sum_{n=\mu}^NK_s^{(n)}\ell_s^{(n)}}\right)\,\exp\left({i\sum_{n=\mu}^NK_i^{(n)}\ell_i^{(n)}}\right)}_{\equiv\beta_0^{(\mu)}}\nonumber\\
        &=\left[\exp\left({i\sum_{m=1}^{\mu-1}(k_s+k_i-\Delta k)L}\right)\,\exp\left({i\sum_{n=\mu+1}^N(k_s+k_i)L}\right)\right]\,\beta_0^{(\mu)}\nonumber\\
        &=\exp\left({i\sum_{\mu=1}^{N-1}(k_s+k_i)L}\right)\,\exp\left({-i\sum_{m=1}^{\mu-1}\Delta k L}\right)\,\beta_0^{(\mu)}\nonumber\\
        &=\exp\left({i(N-1)(\Delta k+k_p)L}\right)\,\exp\left({-i(\mu-1)\Delta k L}\right)\,\beta_0^{(\mu)}.
\end{align}

Note that in \eqref{eqnS13}, the crystal dispersion terms $k_j L$ pertain to passive dispersion acquired by photons passing through the crystal in the absence of a nonlinear interaction. This is the reason that the bounds of the sums over these dispersion terms are written to exclude the index $\mu$ itself; the index $\mu$ specifically indexes the crystal in which the nonlinear interaction does take place. The dispersion associated with the nonlinear interaction is given by $\exp\left(i\frac{\Delta k L}{2}\right)$, and this term is already folded into $JSA_0$.

\end{document}